%% file: paper.ltx
\documentclass{amsart}
\usepackage{subcaption, xcolor, hyperref, mathptmx, amssymb}
\usepackage{url, tikz}

\usetikzlibrary{cd,arrows,shapes,intersections}
\tikzset{thick/.style={line width=.4mm}}
\tikzstyle{vect}=[]
\tikzstyle{tensor}=[circle,thick,draw=black,fill=blue!30,minimum size=4mm]
\tikzstyle{double}=[rounded rectangle,thick,draw=black,fill=yellow!30,minimum width=13.5mm,minimum height = 4mm]
\tikzstyle{hugetensor}=[rounded rectangle,thick,draw=black,minimum width=54mm,minimum height = 4mm]

\theoremstyle{definition}
\newtheorem*{problem*}{Problem}

\theoremstyle{definition}

\newcommand{\<}{\langle}
\renewcommand{\>}{\rangle}
\newcommand{\Heff}{{\pazocal{H}_{\rm eff}}}
\newcommand{\hH}{\pazocal{H}}

\DeclareMathAlphabet{\pazocal}{OMS}{zplm}{m}{n}

\DeclareMathOperator{\id}{id}

\DeclareMathOperator*{\argmin}{arg\,min}

\title{Probabilistic Modeling with Matrix Product States}

\author{James Stokes}
\address{Tunnel, New York, NY}
\email{james@tunnel.tech}
\author{John Terilla}
\address{Tunnel, New York, NY}
\email{john@tunnel.tech}

\begin{document}
\maketitle
 
\begin{abstract}
Inspired by the possibility that generative models based on quantum circuits can provide a useful inductive bias for sequence modeling tasks, we propose an efficient training algorithm for a subset of classically simulable quantum circuit models. The gradient-free algorithm, presented as a sequence of exactly solvable effective models, is a modification of the density matrix renormalization group procedure adapted for learning a probability distribution. The conclusion that circuit-based models offer a useful inductive bias for classical datasets is supported by experimental results on the parity learning problem.
\end{abstract}

\section{Introduction}

The possibility of exponential speedups for certain linear algebra operations has inspired a wave of research into quantum algorithms for machine learning purposes \cite{biamonte2017quantum}.  Many of these exponential speedups hinge on assumptions of fault tolerant quantum devices and efficient data preparation, which are unlikely to be realized in the near future.  Focus has thus shifted to hybrid quantum-classical algorithms which involve optimizing the parameters of a variational quantum circuit to prepare a desired quantum state and have the potential to be implemented on near-term intermediate scale quantum devices \cite{preskill2018quantum}.

Hybrid quantum-classical algorithms have been found to solve difficult eigenvalue problems \cite{peruzzo2014variational} and to perform hard combinatorial optimization \cite{farhi2014quantum}.  A number of recent works consider unsupervised learning within the hybrid quantum-classical framework \cite{han2018unsupervised, cheng2019tree, huggins2018towards, liu2018differentiable, benedetti2018generative, du2018expressive, killoran2018continuous}.

In the context of machine learning, as emphasized in \cite{preskill2018quantum}, it is less clear if variational hybrid quantum-classical algorithms offer advantages over existing purely classical algorithms.  Density estimation, which attempts to learn a probability distribution from training data, has been suggested as an area to look for advantages \cite{benedetti2018generative} because a quantum advantage has been identified in the ability of quantum circuits to sample from certain probability distributions that are hard to sample classically \cite{shepher2009iqc}.  In high-dimensional density estimation relevant to machine learning, expressive power is only part of the story and indeed algorithms in high-dimensional regime rely crucially on their inductive bias.  
Do the highly expressive probability distributions implied by quantum circuits offer a useful inductive bias for modeling high-dimensional classical data?  We address this question in this paper.

 We work within the confines of a classically tractable subset of quantum states modeled by tensor networks, which can be thought of as those states that can be prepared by shallow quantum circuits.  Mathematically, tensor networks are a graphical calculus for describing interrelated matrix factorizations for which there exist polylogarithmic algorithms for a restricted set of linear algebra computations.  We propose an unsupervised training algorithm for a generative model inspired by the density matrix renormalization group (DMRG) procedure.  The training dynamics take place on the unit sphere of a Hilbert space, where in contrast to many variational methods, a state is modified in a sequence of deterministic steps that do not involve gradients. The efficient access to certain vector operations afforded by the tensor network ansatz allows us to implement our algorithm in a purely classical fashion.  

We experimentally probe the inductive bias of the model by training on the dataset $P_{20}$ consisting of bitstrings of length $20$ having an even number of $1$ bits.  The algorithm rapidly learns the uniform distribution on $P_{20}$ to high precision, indicating that the tensor network quantum circuit model provides a useful inductive bias for this classical dataset, which can be frustrating to learn for other models, such as RBMs.  

In an effort to improve accessibility, we avoid the language of quantum-many body physics and quantum information and explain the algorithm and results in terms of elementary linear algebra and statistics.  While this means some motivational material is omitted, we believe it sharpens the exposition.  One exception is the visual language of tensor networks where the benefits of simplifying tensor contractions outweigh the costs of using elementary, but cumbersome, notation.  We refer readers unfamiliar with tensor network notation to \cite{TNORG, schollwock2011density, bridgeman2017hand, orus2014practical} or to the many other surveys.  

The organization of the paper is as follows.  In section \ref{theproblemsection} we state the optimization problem at the population level and propose a finite-sample estimator.  In sections \ref{outlinesection} and \ref{sec:effective} we describe an abstract discrete-time dynamical system evolving on the unit sphere of Hilbert space which optimizes our empirical objective by exactly solving an effective problem in a sequence of isometrically embedded Hilbert subspaces. In section \ref{sec:dmrg} we provide a concrete realization of this dynamical system for a class of tensor networks called matrix product states. Section \ref{sec:experiment} outlines experiments demonstrating that the proposed iterative solver successfully the learns parity language using limited data.

\subsection*{Acknowledgements}  The authors thank Tai-Danae Bradley, Giuseppe Carleo, Joseph Hirsh, Maxim Kontsevich, Miles Stoudenmire, Jackie Shadlen, and Yiannis Vlassopoulos for many helpful conversations.

\section{The problem formulation}\label{theproblemsection}

Recall that a unit vector $\psi$ in a finite-dimensional Hilbert space $\hH$ defines a probability distribution $P_\psi$ on any orthonormal basis by setting the probability of each basis vector $e$ to be
\begin{equation}\label{born}P_\psi(e):= |\<\psi , e\>|^2  .\end{equation}
We refer to the probability distribution $P_\psi$ in Equation \eqref{born} as the \emph{Born distribution} induced by $\psi$.  

Let $\pi$ be a probability distribution on a finite set $\pazocal{X}$ and fix a field of scalars, either $\mathbb{R}$ or $\mathbb{C}$.  Let $\pazocal{H}$ be the free vector space on the set $\pazocal{X}$.  Use $|x\>$ to denote the vector in $\hH$ corresponding to the element $x\in \pazocal{X}$.  The space $\hH$ has a natural inner product defined by declaring the vectors $\{|x\>:x\in \pazocal{X}\}$ to be an orthonormal basis.

Define a unit vector $\psi_\pi \in \pazocal{H}$ by
\begin{equation}\label{defofpsidata}
\psi_\pi := \sum_{x\in \pazocal{X}} \sqrt{\pi(x)} \, |x\>  . 
\end{equation}
Notice that $\psi_\pi$ realizes $\pi$ as a Born distribution:
\begin{equation}
\pi(x)=P_{\psi_\pi}\left(|x\>\right) \text{ for all $x\in \pazocal{X}.$}
\end{equation}  
The formula for $\psi_\pi$ as written in Equation \eqref{defofpsidata} involves perfect knowledge of $\pi$ and unrestricted access to the Hilbert space $\hH$.  This paper is concerned with situations when knowledge about $\pi$ is limited to a finite number of training examples, and $\psi$ is restricted to some tractable subset $\pazocal{M}$ of the unit sphere. 

At the population level, the problem to be solved is to find the closest approximation $\psi_\ast$ to $\psi_\pi$ within $\pazocal{M}$,
\[
\psi_\ast := \argmin_{\psi \in \pazocal{M}} 
\left\| \psi-\psi_\pi\right\|.
\]
We assume access to a sequence $(X_i)_{i=1}^n$ of samples drawn independently from $\pi$, giving rise to the associated empirical distribution
\begin{equation}
\widehat{\pi}(x):=\frac{1}{n}\sum_{i=1}^n \delta_{X_i}(x)  .
\end{equation}
It is natural to define the following estimator whose Born distribution coincides with the empirical distribution
\begin{equation}
	\psi_{\widehat{\pi}} = \sum_{x\in \pazocal{X}} \sqrt{\widehat{\pi}(x)} \, | x \rangle .
\end{equation}
We are thus led to consider the following optimization problem.
\begin{problem*}
Given a sequence $\{X_i\}_{i=1}^n$ of i.i.d.~samples drawn from $\pi$ and a subset $\pazocal{M} \subseteq \ \{ \psi \in \pazocal{H} : \Vert \psi \Vert = 1 \}$ of the unit sphere in $\pazocal{H}$, find
\[
\widehat{\psi} := \argmin_{\psi \in \pazocal{M}} 
\left\| \psi-\psi_{\widehat{\pi}} \right\|.
\]
\end{problem*}
In the case where $\pazocal{X}$ consists of strings, the associated Hilbert space $\pazocal{H}$ has a dimension that is exponential in the string length.  The model hypothesis class $\pazocal{M}$ should be chosen so that the induced Born distribution $P_{\widehat{\psi}}$ offers a useful inductive bias for modeling high-dimensional probability distributions over the space of sequences. We note, parenthetically, that the plug-in estimator $\Vert \psi - \psi_{\widehat{\pi}} \Vert$ is a biased estimator of the population objective $\Vert \psi - \psi_{\pi} \Vert$.

\section{Outline of our approach to solving the problem}\label{outlinesection}
We present an algorithm that, given a fixed realization of data $(x_1,\ldots,x_n) \in \pazocal{X}^n$ and an initial state $\psi_0\in \pazocal{M}$, produces a deterministic sequence $\{\psi_t\}_{t \geq 0}$ of unit vectors in $\pazocal{M}$.
The algorithm is a variation of the density matrix renormalization group (DMRG) procedure which we call \emph{exact single-site DMRG} in which each step produces a vector closer to $\psi_{\widehat{\pi}}$.  The sequence is defined inductively as follows: given $\psi_t$, the inductive step defines a subspace $\hH_{t+1}$ of $\hH$, which also contains $\psi_t$.
Then $\psi_{t+1}$ is defined to be the vector in $\hH_{t+1}$ closest to $\psi_{\widehat{\pi}}$.  Inspired by ideas from the Renormalization Group we provide an analytic formula for $\psi_{t+1}$. The fact that the distance to the target vector $\psi_{\widehat{\pi}}$ decreases after each iteration follows as a simple consequence of the following facts
\begin{equation}
	\psi_t \in \hH_{t+1} \qquad \text{ and }\qquad 
	 \psi_{t+1} = \argmin_{ \{\psi \in \hH_{t+1}:\ \Vert \psi \Vert =1\}}\| \psi_{\widehat{\pi}} - \psi \|.
\end{equation}
See Figure \ref{spheres}.

\input{spheres.tex}

\section{Effective versions of the problem}\label{sec:effective}
Each proposal subspace $\hH_t$ mentioned in the previous section will be defined as the image of an ``effective'' space.  We begin with a general description of an effective space.

Let $\alpha :\Heff \to \hH$ be an isometric embedding of a Hilbert space $\Heff$ into $\hH$.  We refer to $\Heff$ as the effective Hilbert space.  The isometry $\alpha$ and its adjoint map $\alpha^\ast$ are summarized by the following diagram,
\[
\begin{tikzcd}
      \Heff \arrow[rr, bend left, "\alpha"] 
    \arrow[loop,out=210,in=150,distance=20,"\id_{\Heff}"]
    && \pazocal{H} \arrow[ll, bend left,"\alpha^*"]
    \arrow[loop,out=-30,in=30,distance=20,swap,"P"]
\end{tikzcd}
\]
The composition 
$\alpha ^*  \alpha =\id{_\Heff}$ is the identity on $\Heff$.  The composition in the other order $\alpha  \alpha^*$ is an orthogonal projection onto $\alpha(\Heff)$ which is a subspace of $\hH$ isometrically isomorphic to $\Heff$.  Call this orthogonal projection $P$
\begin{equation}
P:=\alpha  \alpha^\ast  .
\end{equation}

The effective version of the problem formulated in Section \ref{theproblemsection} is to find the unit vector $\psi \in \alpha(\Heff)$ in the image of the effective Hilbert space that is closest to $\psi_{\widehat{\pi}}$.  This effective problem is solved exactly in two simple steps.  
The first step is orthogonal projection:  $P( \psi_{\widehat{\pi}})$ is the vector in $\alpha(\Heff)$ closest to $\psi_{\widehat{\pi}}$.   
The second step is to normalize $P( \psi_{\widehat{\pi}})$,
which may not be a unit vector, to obtain the unit vector in $\alpha(\Heff)$ closest to $\psi_{\widehat{\pi}}$.

Therefore, the analytic solution to the effective problem is $P(\psi_{\widehat{\pi}}) / \| P(\psi_{\widehat{\pi}}) \|$ where 
\begin{align}
P(\psi_{\widehat{\pi}}) & = \alpha  \alpha^* \left(\psi_{\widehat{\pi}}\right)  \\
&= \alpha  \alpha^* \left( \sum_{x\in \pazocal{X}} \sqrt{\widehat{\pi}(x)} \, |x\>\right)  \\
&=\alpha \left(\sum_{x \in \pazocal{X}} \sqrt{\widehat{\pi}(x)}\, \alpha^*(|x \>) \right)  .\label{effectivesolution}
\end{align}
In the exact single-site DMRG algorithm, the space $\alpha(\Heff)$ is contained within our model hypothesis class $\pazocal{M}$.  We also offer a multi-site DMRG algorithm in the appendix.  In this multi-site algorithm, the analytic solution to the effective problem in $\alpha(\Heff)$ does not lie in $\pazocal{M}$ so the solution to the effective problem needs to undergo an additional ``model repair'' step.

Before going on to the details of the algorithm, it might be helpful to look more closely at the solution to the effective problem.  
For each training example $x_i$, call the vector $\alpha^\ast(|x_i\>)\in\Heff$ an \emph{effective data point}.  Then, the argument of $\alpha$ in \eqref{effectivesolution} becomes the weighted sum of effective data
\begin{equation}\label{effectivepsi}
\sum_{x\in \pazocal{X}} \sqrt{\widehat{\pi}(x)}\, \alpha^*(|x \>).
\end{equation}
The effective data are not necessarily mutually orthogonal and so the vector in \eqref{effectivepsi} will not be a unit vector.  One may normalize to obtain a unit vector in $\Heff$ and then apply $\alpha$ to obtain the analytic solution to the effective problem.  Normalizing in $\Heff$ and then applying $\alpha$ is the same as applying $\alpha$ and then normalizing in $\hH$ since $\alpha$ is an isometry.

\section{The exact single-site DMRG algorithm}\label{sec:dmrg}
Now specialize to the case that $\pi$ is a probability distribution on a set $\pazocal{X}$ of sequences.  Suppose that $\pazocal{X}=A^N$ consists of sequences of length $N$ in fixed alphabet $A=\{e_1, \ldots, e_d\}$.  The Hilbert space $\hH$, defined as the free Hilbert space on $\pazocal{X}$, has a natural tensor product structure $V^{\otimes N}$ where $V$ is the free Hilbert space on the alphabet $A$.  We refer to $V$ as the \emph{site space}.  So in this situation, the vectors $\{|e_1\>, \ldots, |e_d\>\}$ are an orthonormal basis for the $d$-dimensional site space $V$ and the vectors 
\begin{equation}
|e_{i_1}e_{i_2}\cdots e_{i_N}\> :=|e_{i_1}\> \otimes |e_{i_2}\> \otimes \cdots \otimes |e_{i_N}\>
\end{equation}
are an orthonormal basis for the $d^N$ dimensional space $\hH=V^{\otimes N}$.  
We choose as model hypothesis class the subset $\pazocal{M} \subseteq \pazocal{H}$ consisting of normalized elements in $\hH$ that have a low rank matrix product state (MPS) factorization.  Vectors in this model hypothesis class have efficient representations, even in cases where the Hilbert space $\hH$ is of exponentially high dimension.  
For simplicity of presentation, we consider matrix product states with a single fixed bond space $W$, although everything that follows could be adapted to work with tensor networks without loops having arbitrary bond spaces.   

The exact single-site DMRG algorithm begins with an initial vector $\psi_0 \in \pazocal{M}$ and produces $\psi_1, \psi_2, \ldots$ inductively by solving an effective problem 
\begin{equation}
\hH_{t+1}:= \alpha_{t+1}(\hH_{\mathrm{eff},t+1})
\end{equation}  
which we now describe.  Let us drop the subscript $t+1$ from the isometry $\alpha_{t+1}$ and the effective Hilbert space $\hH_{\mathrm{eff},t+1}$ in the relevant effective problem---just be aware that the embedding
\begin{equation}
\alpha:\Heff \to \hH
\end{equation}
will change from step to step.  The map $\alpha$ is defined using an MPS factorization  of $\psi_{t}$ in mixed canonical form relative to a fixed site which varies at each step according to a predetermined schedule.  For the purposes of illustration, the third site is the fixed site in the pictures below.
\begin{equation}
    \begin{tikzpicture}[x=0.75cm,y=0.75cm,baseline={(current bounding box.center)}]
        \node[] (psi) at (-1,0) {$\psi_t=$};
        \node[tensor] (m0) at (0,0) {};
        \node[tensor] (m1) at (1,0) {};
        \node[tensor] (m2) at (2,0) {};
        \node[tensor] (m3) at (3,0) {};
        \node[tensor] (m4) at (4,0) {};
        \node[tensor] (m5) at (5,0) {};
        \node[tensor] (m6) at (6,0) {};
        \node[] (i0) at (0,1) {};
        \node[] (i1) at (1,1) {};
        \node[] (i2) at (2,1) {};
        \node[] (i3) at (3,1) {};
        \node[] (i4) at (4,1) {};
        \node[] (i5) at (5,1) {};
        \node[] (i6) at (6,1) {};

        \node[circle,thick, dashed,draw=black,fill=none,minimum size=7mm] at (2,0) {};

        \draw [thick] (i0) -- (m0);
        \draw [thick] (i1) -- (m1);
        \draw [thick] (i2) -- (m2);
        \draw [thick] (i3) -- (m3);
        \draw [thick] (i4) -- (m4);
        \draw [thick] (i5) -- (m5);
        \draw [thick] (i6) -- (m6);
               
        \draw [thick] (m0) -- (m1) -- (m2) -- (m3) -- (m4) -- (m5) -- (m6);
\end{tikzpicture}
\end{equation}
The effective space is $\Heff=W\otimes V\otimes W$ and the isometric embedding $\alpha:W\otimes V\otimes W \to V^{\otimes N}$ is defined for any $\phi \in W\otimes V\otimes W$ by replacing the tensor at the fixed site of 
$\psi_t$ with $\phi$:
\begin{equation}
   \begin{tikzpicture}[x=0.75cm,y=0.75cm,baseline={(current bounding box.center)}]
        \node[tensor, fill=white] (d1) at (-3,0) {}; 
        \node[] (top1) at (-3,1) {}; 
        \node[] (left1) at (-4,0) {};
        \node[] (right1) at (-2,0) {};  
        \node[] (d3) at (-2,0) {}; 
        \node[] (d4) at (-.5,0) {}; 
        
        \draw [thick] (left1) -- (d1) -- (right1);
        \draw [thick] (top1) -- (d1);

        \draw[thick, |->]  (d3) -- node[above,yshift=4pt] {$\alpha$} (d4) ;

        \node[tensor] (m0) at (0,0) {};
        \node[tensor] (m1) at (1,0) {};
        \node[tensor, fill=white] (m2) at (2,0) {};
        \node[tensor] (m3) at (3,0) {};
        \node[tensor] (m4) at (4,0) {};
        \node[tensor] (m5) at (5,0) {};
        \node[tensor] (m6) at (6,0) {};

        \node[] (i0) at (0,1) {};
        \node[] (i1) at (1,1) {};
        \node[] (i2) at (2,1) {};
        \node[] (i3) at (3,1) {};
        \node[] (i4) at (4,1) {};
        \node[] (i5) at (5,1) {};
        \node[] (i6) at (6,1) {};

        \draw [thick] (i0) -- (m0);
        \draw [thick] (i1) -- (m1);
        \draw [thick] (i2) -- (m2);
        \draw [thick] (i3) -- (m3);
        \draw [thick] (i4) -- (m4);
        \draw [thick] (i5) -- (m5);
        \draw [thick] (i6) -- (m6);
               
        \draw [thick] (m0) -- (m1) -- (m2) -- (m3) -- (m4) -- (m5) -- (m6);
        \end{tikzpicture}
\end{equation}
To see that $\alpha$ is an isometry, use the gauge condition that the MPS factorization of $\psi_{t}$ is in mixed canonical form relative to the fixed site, as illustrated below:
\begin{equation}
    \begin{tikzpicture}[x=0.75cm,y=0.75cm,baseline={(current bounding box.center)}]

        \node[] (psi) at (-2,1) {$\<\alpha(\phi), \alpha(\phi')\>=$};
        
        \node[tensor] (n0) at (0,2) {};
        \node[tensor] (n1) at (1,2) {};
        \node[tensor, fill=white] (n2) at (2,2) {};
        \node[tensor] (n3) at (3,2) {};
        \node[tensor] (n4) at (4,2) {};
        \node[tensor] (n5) at (5,2) {};

        \node[tensor] (m0) at (0,0) {};
        \node[tensor] (m1) at (1,0) {};
        \node[tensor, fill=green!25] (m2) at (2,0) {};
        \node[tensor] (m3) at (3,0) {};
        \node[tensor] (m4) at (4,0) {};
        \node[tensor] (m5) at (5,0) {};

        \node[] (i0) at (0,1) {};
        \node[] (i1) at (1,1) {};
        \node[] (i2) at (2,1) {};
        \node[] (i3) at (3,1) {};
        \node[] (i4) at (4,1) {};
        \node[] (i5) at (5,1) {};

        \draw [thick] (n0) -- (m0);
        \draw [thick] (n1) -- (m1);
        \draw [thick] (n2) -- (m2);
        \draw [thick] (n3) -- (m3);
        \draw [thick] (n4) -- (m4);
        \draw [thick] (n5) -- (m5);

        \draw [thick] (n0) -- (n1) -- (n2) -- (n3) --(n4) --(n5);
        \draw [thick] (m0) -- (m1) -- (m2) -- (m3) --(m4) --(m5);

        \node[] (psi) at (5.5,1) {$=$};
        
        \node[tensor,fill=green!25] (mp) at (7,0) {};
        \node[tensor, fill=white] (np) at (7,2) {};
        \draw [thick,] (mp)  -- (np);
        \draw[thick,] (np) to[out=180,in=180] (mp);            
        \draw[thick,] (np) to[out=0,in=0] (mp);    

        \node[] (psi) at (9,1) {$=\<\phi, \phi'\>.$};        
\end{tikzpicture}
\end{equation}
The adjoint map $\alpha^*:V^{\otimes N} \to W\otimes V \otimes W$ has a clean pictorial depiction as well.
\begin{equation}
    \begin{tikzpicture}[x=0.75cm,y=0.75cm,baseline={(current bounding box.center)}]
        \node[hugetensor,fill=red!10,minimum width=48mm] (v17) at (-5.5,1) {};
        \node[] (j1) at (-3,1) {};
        \node[] (j2) at (-4,1) {};
        \node[] (j3) at (-5,1) {};
        \node[] (j4) at (-6,1) {};
        \node[] (j5) at (-7,1) {};
        \node[] (j6) at (-8,1) {};
        \node[] (n1) at (-3,0) {};
        \node[] (n2) at (-4,0) {};
        \node[] (n3) at (-5,0) {};
        \node[] (n4) at (-6,0) {};
        \node[] (n5) at (-7,0) {};
        \node[] (n6) at (-8,0) {};

        \draw [thick,shorten <=0.7mm] (j1) -- (n1);
        \draw [thick,shorten <=0.7mm] (j2) -- (n2);
        \draw [thick,shorten <=0.7mm] (j3) -- (n3);
    
        \draw [thick,shorten <=0.7mm] (j4) -- (n4);
        \draw [thick,shorten <=0.7mm] (j5) -- (n5);
        \draw [thick,shorten <=0.7mm] (j6) -- (n6);
               
        \node[] (d3) at (-2,.5) {}; 
        \node[] (d4) at (0,.5) {}; 
        \draw[thick, |->]  (d3) -- node[above,yshift=4pt] {$\alpha^*$} (d4) ;

        \node[tensor] (m0) at (1,0) {};
        \node[tensor] (m1) at (2,0) {};

        \node[] (m2) at (3,0) {};
        \node[tensor] (m3) at (4,0) {}; 

        \node[tensor] (m4) at (5,0) {};
        \node[tensor] (m5) at (6,0) {};

        \node[hugetensor,fill=red!10,minimum width=48mm] (v17) at (3.5,1) {};
        \node[vect] (i0) at (1,1) {};
        \node[vect] (i1) at (2,1) {};
        \node[vect] (i2) at (3,1) {};
        \node[vect] (i3) at (4,1) {};
        \node[vect] (i4) at (5,1) {};
        \node[vect] (i5) at (6,1) {};

        \draw [thick,shorten <=0.7mm]  (i1) -- (m1);
        \draw [thick,shorten <=0.7mm] (i2) -- (m2);
        \draw [thick,shorten <=0.7mm] (i3) -- (m3);

        \draw [thick,shorten <=0.7mm] (i0) -- (m0);
    
        \draw [thick,shorten <=0.7mm] (i4) -- (m4);
        \draw [thick,shorten <=0.7mm] (i5) -- (m5);
               
       \draw [thick] (m0) -- (m1) -- (m2) ;
       \draw [thick] (m2) -- (m3) -- (m4) -- (m5) -- (m6);
\end{tikzpicture}
\end{equation}
To see that $\alpha^*$ as pictured above is, in fact, the adjoint of $\alpha$, note that for any $\eta \in \hH$ and any $\phi \in \Heff$, both $\<\eta, \alpha(\phi)\>$ and $\<\alpha^*(\eta),\phi\>$ result in the same tensor contraction:
\begin{equation}
\begin{tikzpicture}[x=0.75cm,y=0.75cm,baseline={(current bounding box.center)}]

        \node[] (psi) at (-2,1) {$\<\eta, \alpha(\phi)\>=$};
        
        \node[hugetensor,minimum width=48mm, fill=red!10] (v17) at (2.5,2) {};
        \node[] (n0) at (0,2) {};
        \node[] (n1) at (1,2) {};
        \node[] (n2) at (2,2) {};
        \node[] (n3) at (3,2) {};
        \node[] (n4) at (4,2) {};
        \node[] (n5) at (5,2) {};

        \node[tensor] (m0) at (0,0) {};
        \node[tensor] (m1) at (1,0) {};
        \node[tensor, fill=yellow!25] (m2) at (2,0) {};
        \node[tensor] (m3) at (3,0) {};
        \node[tensor] (m4) at (4,0) {};
        \node[tensor] (m5) at (5,0) {};

        \node[] (i0) at (0,1) {};
        \node[] (i1) at (1,1) {};
        \node[] (i2) at (2,1) {};
        \node[] (i3) at (3,1) {};
        \node[] (i4) at (4,1) {};
        \node[] (i5) at (5,1) {};

        \draw [thick, shorten <=0.7mm] (n0) -- (m0);
        \draw [thick, shorten <=0.7mm] (n1) -- (m1);
        \draw [thick, shorten <=0.7mm] (n2) -- (m2);
        \draw [thick, shorten <=0.7mm] (n3) -- (m3);
        \draw [thick, shorten <=0.7mm] (n4) -- (m4);
        \draw [thick, shorten <=0.7mm] (n5) -- (m5);

        \draw [thick] (m0) -- (m1) -- (m2) -- (m3) -- (m4) -- (m5);
     
        \node[] (psi) at (7,1) {$=\<\alpha^*(\eta), \phi\>$};        
\end{tikzpicture}
\end{equation}
In the picture above, begin with the blue tensors.  Contracting with the yellow tensor gives $\alpha(\phi)$ and then contracting with the red tensor gives $\< \eta, \alpha(\phi)\>$.  On the other hand, first contracting with the red tensor yields $\alpha^*(\eta)$ resulting in $\<\alpha^*(\eta), \phi\>$ after contracting with the yellow tensor.

Now, Equation \eqref{effectivesolution} describes an analytic solution for the vector in 
$\hH_{t+1} := \alpha(W\otimes V\otimes W)$ closest to $\psi_{\widehat{\pi}}$.  Namely, $\alpha(\phi/\|\phi\|)$ where 
\begin{equation}\label{effectivedata}
\phi = \sum_{x \in \pazocal{X}}  \sqrt{\widehat{\pi}(x)} \, \alpha^*(|x\>).
\end{equation}
For each sample $|x_i\rangle = |e_{i_1}e_{i_2}\cdots e_{i_N}\rangle $, the effective data point $\alpha^\ast(|x_i\>)\in V \otimes W \otimes V$ is given by the contraction
\begin{equation}\label{effectivedata}
\begin{tikzpicture}[x=0.75cm,y=0.75cm,baseline={(current bounding box.center)}]
        \node[] () at (-1.5,0) {$\alpha^*\left(|x_i\rangle\right)=$};
        \node[tensor] (m0) at (0,0) {};
        \node[tensor] (m1) at (1,0) {};
        \node[] (m2) at (2,0) {};
        \node[tensor] (m3) at (3,0) {};
        \node[tensor] (m4) at (4,0) {};
        \node[tensor] (m5) at (5,0) {};
        \node[tensor] (m6) at (6,0) {};

        \node[] () at (6.75,0) {$=$};
        \node[tensor, fill=yellow!10] (Xi) at (8,0) {};
        \node[] (b0) at (7,0) {};
        \node[] (b1) at (9,0) {};
        \node[] (v1) at (8,1) {};
        \draw [thick] (b0) -- (Xi) -- (b1);
        \draw [thick] (v1) -- (Xi);

        \node[] (i0) at (0,1) {$e_{i_1}$};
        \node[] (i1) at (1,1) {$e_{i_2}$};
        \node[] (i2) at (2,1) {$e_{i_3}$};
        \node[] (i3) at (3,1) {$e_{i_4}$};
        \node[] (i4) at (4,1) {$e_{i_5}$};
        \node[] (i5) at (5,1) {$e_{i_6}$};
        \node[] (i6) at (6,1) {$e_{i_7}$};

        \draw [thick] (i0) -- (m0);
        \draw [thick] (i1) -- (m1);
        \draw [thick] (i2) -- (m2);
        \draw [thick] (i3) -- (m3);
        \draw [thick] (i4) -- (m4);
        \draw [thick] (i5) -- (m5);
        \draw [thick] (i6) -- (m6);
               
        \draw [thick] (m0) -- (m1) -- (m2) -- (m3) -- (m4) -- (m5) -- (m6);
\end{tikzpicture}
\end{equation}
Once the effective form $\alpha^\ast(|x\>)$ of each distinct training example $| x \rangle$ has been computed, weighted by $\sqrt{\widehat{\pi}(x)}$, summed, and normalized, one obtains an expression for the unit vector $\phi / \| \phi \| \in W\otimes V \otimes W$, depicted as follows,
\begin{equation}\label{updatedtensor}
\begin{tikzpicture}[x=0.75cm,y=0.75cm,baseline={(current bounding box.center)}]
        \node[] () at (-4,.25) {$\dfrac{\phi}{\| \phi\|}=$};
        \node[tensor, fill=yellow!100] (i) at (-2,0) {};
        \node[] (m0) at (0,0) {};
 
        \node[] (v0) at (-3,0) {};
        \node[] (v1) at (-1,0) {};
        \node[] (b1) at (-2,1) {};
        \draw [thick] (v0) -- (i) -- (v1);
        \draw [thick] (b1) -- (i) ;
\end{tikzpicture}
\end{equation}
Finally, apply the map $\alpha$ to get $\psi_{t+1}$:
\begin{equation}\label{updatedMPS}
    \begin{tikzpicture}[x=0.75cm,y=0.75cm,baseline={(current bounding box.center)}]
        \node[] (psi) at (-1.5,0) {$\psi_{t+1}=$};
        \node[tensor] (m0) at (0,0) {};
        \node[tensor] (m1) at (1,0) {};
        \node[tensor, fill=yellow!100] (m2) at (2,0) {};
        \node[tensor] (m3) at (3,0) {};
        \node[tensor] (m4) at (4,0) {};
        \node[tensor] (m5) at (5,0) {};
        \node[tensor] (m6) at (6,0) {};

        \node[] (i0) at (0,1) {};
        \node[] (i1) at (1,1) {};
        \node[] (i2) at (2,1) {};
        \node[] (i3) at (3,1) {};
        \node[] (i4) at (4,1) {};
        \node[] (i5) at (5,1) {};
        \node[] (i6) at (6,1) {};

        \draw [thick] (i0) -- (m0);
        \draw [thick] (i1) -- (m1);
        \draw [thick] (i2) -- (m2);
        \draw [thick] (i3) -- (m3);
        \draw [thick] (i4) -- (m4);
        \draw [thick] (i5) -- (m5);
        \draw [thick] (i6) -- (m6);
               
        \draw [thick] (m0) -- (m1) -- (m2) -- (m3) -- (m4) -- (m5) -- (m6);
\end{tikzpicture}
\end{equation}
To complete the description of the exact single-site DMRG algorithm, we need to choose a schedule in which to update the tensors.  We use the following schedule, organized into back-and-forth sweeps, for the fixed site at each step
\begin{equation}
\underbrace{1,2, 3, \ldots, N-1, N, N-1, \ldots, 3, 2,}_{\text{ Sweep 1}} \underbrace{1, 2, \ldots, N-1, N, N-1, \ldots, 2,}_{\text{ Sweep 2}}1, 2, \ldots 
\end{equation}

A schedule that proceeds by moving the fixed site one position at a time allows us to take advantage of two efficiencies resulting in an algorithm that is linear in both the number of training examples $n$ and the number of sites $N$.  One efficiency is that most of the calculations of the effective data in Equation \eqref{effectivedata} used to compute $\psi_{t+1}$ can be reused when computing $\psi_{t+2}$.  The second efficiency is that when inserting the updated tensor in Equation \eqref{updatedtensor}, it can be done so that the resulting MPS factorization of $\psi_{t+1}$ as pictured in Equation \eqref{updatedMPS} will be in mixed canonical form relative to a site adjacent to the updated tensor, which avoids a costly gauge fixing step.

\section{Experiments}\label{sec:experiment}
\begin{figure}
\includegraphics[width=0.8\textwidth]{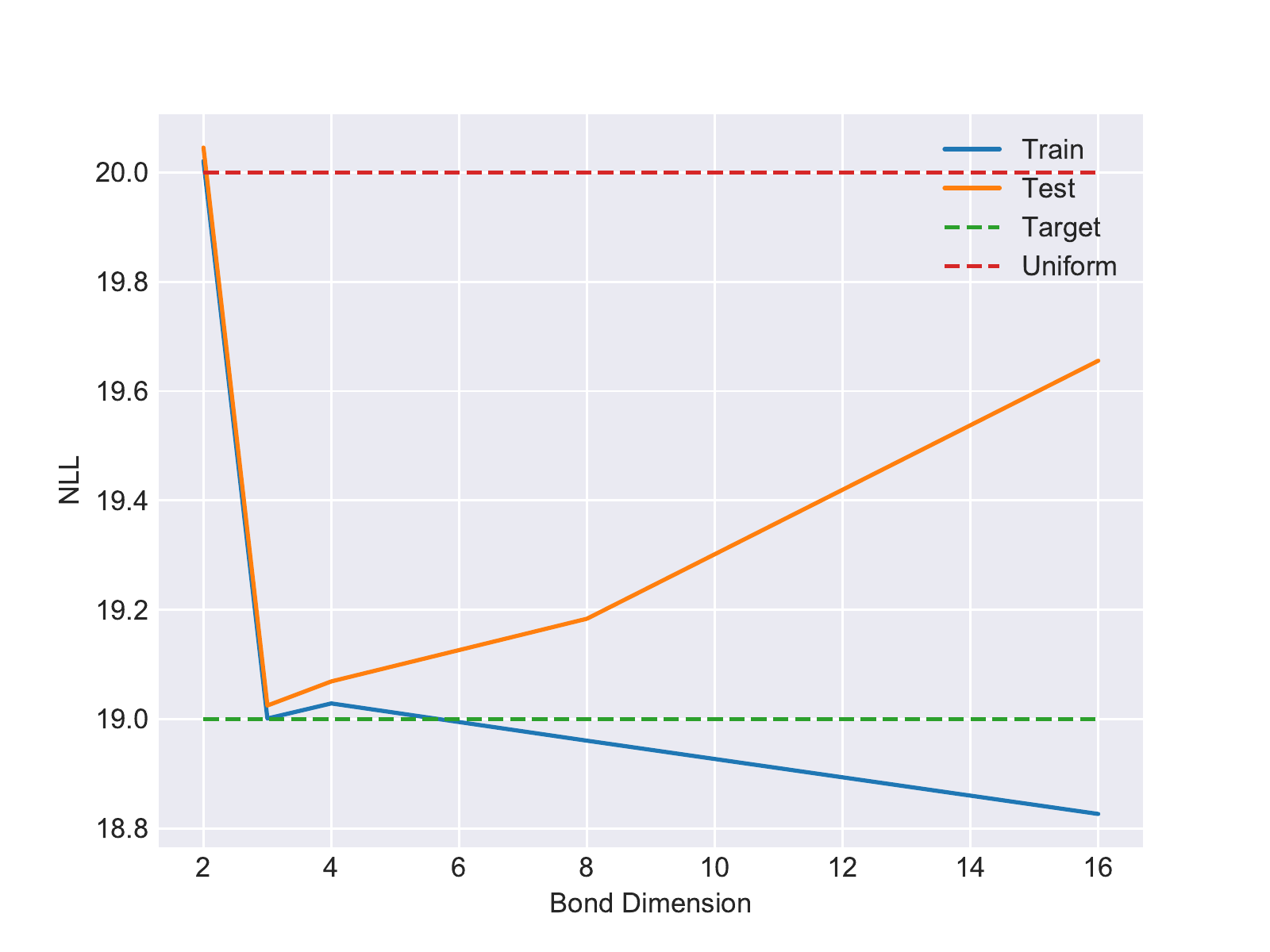}
\caption{A representative bias-variance tradeoff curve showing negative log-likelihood (base 2) as a function of bond dimension for exact single-site DMRG on the $P_{20}$ dataset.  For bond dimension $3$, the generalization gap is approximately $\epsilon = 0.0237$.  For reference, the uniform distribution on bitstrings has NLL of $20$.  Memorizing the training data would yield a NLL of approximately $13.356$.
\label{fig:P20}}

\end{figure}

This section considers the problem of unsupervised learning of the parity language $P_N$, which consists of bitstrings of length $N$ containing an even number of 1 bits. Of the total $2^{N-1}$ such bitstrings, we reserved random disjoint subsets of size $2\%$ for training, cross-validation and testing purposes. A NLL of $N-1$ corresponds to the entropy of the uniform distribution on $P_N$.  If the model memorizes the training set, it will assign to it a negative-log-likelihood (NLL) of $N-1+\log_2(0.02)$ corresponding to the entropy of the uniform distribution on the training data.  A NLL of $N$ corresponds to the entropy of the uniform distribution on all bitstrings of length $N$.  The measure of generalization performance is the gap $\epsilon$ between the NLL of the training and testing data.  We performed exact single-site DMRG over the real number field using the $P_{20}$ dataset for different choices of bond dimension.  Training was terminated according to an early stopping criterion as determined by distance between the MPS state and the state of the cross-validation sample.  The NLL as a function of bond dimension reported in Fig.~\ref{fig:P20} displays the expected bias-variance tradeoff, with optimal model complexity occurring at bond dimension $3$ with corresponding generalization gap $\epsilon = 0.0237$.  

\section{Discussion}
A number of recent works have explored the parity dataset using restricted Boltzmann machines (RBMs) and found it to be difficult to learn, even in experiments that train using the entire dataset \cite{Romero2018WCD, Montufar2011EPA}.  Recall that an RBM is a universal approximator of distributions on $\{0,1\}^N$, given sufficiently many hidden units.  Ref.~\cite{Montufar2011EPA} proved that any probability distribution on $\{0,1\}^N$ can be approximated within $\epsilon$ in KL-divergence by an RBM with  
$
m \geq 2^{(N-1)(1-\epsilon) + 0.1}
$
hidden units.  For $P_{20}$ this bound works out to be about $4 \times 10^5$ hidden nodes.  It would be interesting to know whether it could be learned with significantly fewer.

It is not difficult to train a feedforward neural network to classify bitstrings by parity using labelled data, but we do not know if there are unsupervised generative neural models that do well learning $P_N$.  It is reasonable to expect that recurrent models whose training involve conditional probabilities $\pi(x_1, \ldots, x_k | x_{k+1}, \ldots, x_N)$ might be frustrated by $P_N$ since the conditional distributions contain no information:  any bitstring of length less than $N$ has the same number of completions in $P_N$ as not in $P_N$.  

The reader may be interested in \cite{amin2018quantum, kappen2018learning} where quantum models are used to learn classical data.  Those works considered quantum Boltzman machines which were shown to learn the distribution more effectively than their classical counterparts using the same dataset.  The complexity of classically simulating a QBM scales exponentially with the number of sites in contrast to the tensor network algorithms presented here, which scale linearly in the number of sites.

\appendix

\section{Multi-site DMRG}
For completeness we now describe a related \emph{multi-site DMRG} algorithm. The model class $\pazocal{M}$ now consists of normalized vectors with matrix product factorizations, with possibly different bond spaces having dimension less than a fixed upper bound.  The algorithm begins with an initial vector $\psi_0 \in \pazocal{M}$ and produces $\psi_1, \psi_2, \ldots$ inductively.  The inductive step is similar in that we solve an effective problem in the image of an effective Hilbert space
\begin{equation}
\hH_{t+1}:= \alpha_{t+1}(\hH_{\mathrm{eff},t+1})
\end{equation}
to find the unit vector in $\hH_{t+1}$ that is closest to the target state $\psi_{\widehat{\pi}}$, which we now denote with a tilde:
\begin{equation}
\widetilde{\psi}_{t+1}: = \argmin_{ \{\psi \in \hH_{t+1}:\ \Vert \psi \Vert =1\}}\| \psi_{\widehat{\pi}} - \psi \|.
\end{equation}
In multi-site DMRG, as opposed to single-site DMRG, the image of the effective space $\hH_{t+1}$ is not contained in the MPS model hypothesis class $\pazocal{M}$.  So, the solution $\widetilde{\psi}_{t+1}$ to the effective problem must undergo a ``model repair" step
\begin{equation}\label{modelrepair}
\widetilde{\psi}_{t+1} \leadsto \psi_{t+1}
\end{equation}
to produce a vector $\psi_{t+1}\in \pazocal{M}$. In summary:
\begin{itemize}
    \item Use $\psi_t$ to define an isometric embedding $\alpha_{t+1}:\Heff \to \hH$ with $\psi_t \in \hH_{t+1}:=\alpha_{t+1}(\Heff).$
    \item Let $\widetilde{\psi}_{t+1}$ be the unit vector in $\hH_{t+1}$ closest to $\psi_{\widehat{\pi}}$.
    \item Perform a model repair of $\widetilde{\psi}_{t+1}$ to obtain a vector ${\psi}_{t+1} \in \pazocal{M}.$  There are multiple ways to do the model repair.
\end{itemize}

\begin{figure}[!h]
   \begin{tikzpicture}[rotate=70, 
   every node/.style={draw=black, circle, fill, minimum size=2mm, inner sep=0}]
    \node [fill=red!70,label=left :$\psi_{t}$] (A) at (7, 6.8) {}; 
    \node [label=below left :$\psi_{\widehat{\pi}}$] (B) at (7.75, 3.9) {}; 
    \node [fill=red!70,label=right :$\widetilde{\psi}_{t+1}$] (C) at (7.25, 2.5) {};
    \node [fill=red!70,label=below :$\psi_{t+1}^{\rm SVD}$] (D) at (5.6, 1.5) {};
    \node [fill=red!70,label=right :$\psi^{\rm better}_{t+1}$] (E) at (9.6, 3.1) {};

\path [draw=none,fill=blue, fill opacity = 0.1,even odd rule] 
(7,3.5) ellipse (3 and 4) (7.25,2.5) circle (1.92);

\draw [thin, dashed] (A) -- (B) ;
\draw [thin, dashed] (C) -- (B) ;
\draw [thin, dashed] (D) -- (B) ;
\draw [thin, dashed] (E) -- (B) ;
\draw [thin, dashed] (C) -- (D) ;
\draw [thin, dashed] (C) -- (E) ;
              
\end{tikzpicture}

\captionsetup{singlelinecheck=off}
\caption[modelrepair]
{The shaded region represents the model class $\pazocal{M}$.  The red points all lie in $\hH_{t+1}$.
The vector 
$\widetilde{\psi}_{t+1}$ is defined to be the unit vector in $\hH_{t+1}$ closest to the target $\psi_{\widehat{\pi}}$.  Note that 
$\widetilde{\psi}_{t+1}$ does not lie in $\pazocal{M}$.  The vector $\psi_{t+1}^{\rm SVD}$ is defined to be the vector in $\pazocal{M} \cap \hH_{t+1}$  closest to to $\widetilde{\psi}_{t+1}$.  In this picture, $\| \psi_{t+1}^{\rm SVD} - \psi_{\widehat{\pi}} \| > \| \psi_t - \psi_{\widehat{\pi}} \|.$  There may be a point, such as the one labelled 
$\psi_{t+1}^{\rm better}$, which lies in $\pazocal{M}\cap \hH_{t+1}$ and is
closer to $\psi_{\widehat{\pi}}$ than $\psi_{t+1}^{\rm SVD} $, notwithstanding the fact that is is further from $\widetilde{\psi}_{t+1}$.

This figure, to scale, depicts a scenario in which $\| \psi_{t} - \psi_{\widehat{\pi}} \| = .09$, $\| \psi_{t+1}^{\rm SVD} - \psi_{\widehat{\pi}} \|=.10$, $\| \psi_{t+1}^{\rm better} - \psi_{\widehat{\pi}} \|=.07$, $\| \widetilde{\psi}_{t+1} - \psi_{\widehat{\pi}} \|=.06$, $\| \psi_{t+1}^{\rm SVD} - \widetilde{\psi}_{t+1} \| = .07$, and $\| \psi_{t+1}^{\rm better} - \widetilde{\psi}_{t+1} \| = .08.$
  } \label{modelrepair}
\end{figure}

In order to define the effective problem in the inductive step of multi-site DMRG, one uses an MPS factorization of $\psi_t$ in mixed canonical gauge relative to an interval of $r$-sites.  In the picture below, the interval consists of the two sites $3$ and $4$.
\begin{equation}
    \begin{tikzpicture}[x=0.75cm,y=0.75cm,baseline={(current bounding box.center)}]
        \node[] (psi) at (-1,0) {$\psi_t=$};
        \node[tensor] (m0) at (0,0) {};
        \node[tensor] (m1) at (1,0) {};
        \node[tensor] (m2) at (2,0) {};
        \node[tensor] (m3) at (3,0) {};
        \node[tensor] (m4) at (4,0) {};
        \node[tensor] (m5) at (5,0) {};
        \node[tensor] (m6) at (6,0) {};

        \node[rounded rectangle,thick,dashed, draw=black,minimum width=18mm,minimum height = 6mm] at (2.5,0) {};

        \node[] (i0) at (0,1) {};
        \node[] (i1) at (1,1) {};
        \node[] (i2) at (2,1) {};
        \node[] (i3) at (3,1) {};
        \node[] (i4) at (4,1) {};
        \node[] (i5) at (5,1) {};
        \node[] (i6) at (6,1) {};

        \draw [thick] (i0) -- (m0);
        \draw [thick] (i1) -- (m1);
        \draw [thick] (i2) -- (m2);
        \draw [thick] (i3) -- (m3);
        \draw [thick] (i4) -- (m4);
        \draw [thick] (i5) -- (m5);
        \draw [thick] (i6) -- (m6);
               
        \draw [thick] (m0) -- (m1) -- (m2) -- (m3) -- (m4) -- (m5) -- (m6);
\end{tikzpicture}
\end{equation}
The effective Hilbert space $\Heff=W_L \otimes V^{\otimes r} \otimes W_R$ where $W_L$ and $W_R$ are the bond spaces to the left and right of the fixed interval of sites, and $r$ is the length of the chosen interval.  The map $\alpha:W_L \otimes V^{\otimes r} \otimes W_R \to V^{\otimes n}$ is given by replacing the interval of sites and contracting
\begin{equation}
    \begin{tikzpicture}[x=0.75cm,y=0.75cm,baseline={(current bounding box.center)}]

    \node[] (d0) at (-3.5,0) {};  
    \node[double, fill=white] (d1) at (-2,0) {}; 
    \node[] (d2) at (-0.5,0) {};  
 
    \node[] (e0) at (-2.5,1) {};  
    \node[] (e2) at (-1.5,1) {};  
    
    \draw [thick, shorten >=3.7mm]
            (e0) |- (d1.north);
    \draw [thick, shorten >=3.7mm]
            (e2) |- (d1.north);
    \draw [thick] (d0) -- (d1) -- (d2);
        
            \node[] (d3) at (-.25,.25) {}; 
    \node[] (d4) at (2.5,.25) {}; 
        \draw[thick, |->]  (d3) -- node[above,yshift=4pt] {$\alpha$} (d4) ;

        \node[tensor] (m0) at (3,0) {};
        \node[tensor] (m1) at (4,0) {};
         \node[double, fill=white] (m2) at (5.5,0) {}; 
        \node[tensor] (m4) at (7,0) {};
        \node[tensor] (m5) at (8,0) {};
        \node[tensor] (m6) at (9,0) {};

        \node[] (i0) at (3,1) {};
        \node[] (i1) at (4,1) {};
        \node[] (i2) at (5,1) {};
        \node[] (i3) at (6,1) {};
        \node[] (i4) at (7,1) {};
        \node[] (i5) at (8,1) {};
        \node[] (i6) at (9,1) {};

        \draw [thick] (i1) -- (m1);
        \draw [thick, shorten >=3.7mm]
            (i2) |- (m2.north);
        \draw [thick, shorten >=3.7mm]
            (i3) |- (m2.north);

        \draw [thick] (i0) -- (m0);
    
        \draw [thick] (i4) -- (m4);
        \draw [thick] (i5) -- (m5);
        \draw [thick] (i6) -- (m6);
               
       \draw [thick] (m0) -- (m1) -- (m2) -- (m4) -- (m5) -- (m6);
\end{tikzpicture}
\end{equation}
The map $\alpha$ and its adjoint $\alpha^\ast$ are described by, and have properties proved by, pictures completely analogous to those detailed for single-site DMRG in Section \ref{sec:dmrg}.  The effective problem is also solved the same way.  What is not the same is that the vector in $\hH_{t+1} = \alpha(W_L \otimes V^{\otimes r}\otimes W_R)$ which solves the effective problem is \emph{outside} of the model class $\pazocal{M}$ and so one performs a model repair step $\widetilde{\psi}_{t+1} \leadsto \psi_{t+1}$, pictured graphically in $\Heff$ by:
\begin{equation}
    \begin{tikzpicture}[x=0.75cm,y=0.75cm,baseline={(current bounding box.center)}]
    \node[] (d0) at (-3.5,0) {};  
    \node[double, fill=white] (d1) at (-2,0) {}; 
    \node[] (d2) at (-0.5,0) {};  
    \node[] (e0) at (-2.5,1) {};  
    \node[] (e2) at (-1.5,1) {};  
    
    \draw [thick, shorten >=3.7mm]
            (e0) |- (d1.north);
    \draw [thick, shorten >=3.7mm]
            (e2) |- (d1.north);
    \draw [thick] (d0) -- (d1) -- (d2);

        \node[] (d4) at (1.5,.25) {}; 
        \node[scale=2] (4,.25) {$\leadsto$};
        \node[tensor, fill=white] (m0) at (2,0) {};
        \node[tensor, fill=white] (m1) at (3,0) {};
        \node[] (i0) at (2,1) {};
        \node[] (i1) at (3,1) {};
        \node[] (m-1) at (1,0) {};
        \node[] (m3) at (4,0) {};
 \draw [thick] (m-1) -- (m0) -- (m1) -- (m3);    
        \draw [thick] (i0) -- (m0) -- (m1) -- (i1);
     
\end{tikzpicture}
\end{equation}

One way to perform the model repair is to choose
\begin{equation}\label{equation31}
\psi_{t+1}: = \argmin_{\psi \in \pazocal{M}\cap \hH_{t+1}} \| \psi - \widetilde{\psi}_{t+1} \|
\end{equation}  
but the flexibility of the model repair step allows for other possibilities.  One can use the model repair to implement a dynamic tradeoff between proximity to $\widetilde{\psi}_{t+1}$ and other constraints of interest, such as bond dimension.  Many of these implementations have good algorithms arising from singular value decompositions manageable in the effective Hilbert space.  Let use denote such a model repair choice as $\psi_{t+1}^{\rm SVD}$.
Be aware that if  $\psi_{t+1}^{\rm SVD}$ is the vector in $\pazocal{M}\cap \hH_{t+1}$ nearest to $\widetilde{\psi}_{t+1}$ as in Equation \eqref{equation31}, there is no guarantee that $\psi_{t+1}^{\rm SVD}$ will be nearer to $\psi_{\widehat{\pi}}$ than the previous iterate.
In fact, we have experimentally observed the sequence obtained by this kind of model repair to move away from $\psi_{\widehat{\pi}}$.  See Figure \ref{modelrepair} for an illustration of this possibility.  

One might hope to improve the model repair step, say by pre-conditioning the singular value decomposition in a way that is knowledgeable about the target $\psi_{\widehat{\pi}}$.    For the experiments reported in this paper, single-site DMRG consistently outperformed multi-site DMRG for several choices of model repair step, and we include multi-site DMRG only for pedagogical reasons.  The adaptability of the bond dimension afforded by the multi-site DMRG algorithm could provide benefits that outweigh the challenges of good model repair in some situations.

\bibliographystyle{plain}
\bibliography{mps}

\end{document}

%% file: spheres.tex
\begin{figure}[h!]
    \begin{subfigure}[t]{0.45\textwidth}
        \centering
        \resizebox{\linewidth}{!}
        {
        \hspace{12pt}
        \begin{tikzpicture}[every node/.style={draw,circle,minimum size=1.5mm,inner sep=0pt,outer sep=0pt,fill=black}]
            \draw[thick] (0,0) circle (2) ;
            \node [label=left :$\psi_{0}$] at (-1.92,-.55) {};
            \node [label=above :$\psi_{\widehat{\pi}}$] at (1.2,1.6) {};  
        \end{tikzpicture}
        }
        \caption{The initial vector $\psi_0$ and the vector $\psi_{\widehat{\pi}}$ lie in the unit sphere of $\hH$.}
    \end{subfigure}
\hfill
    \begin{subfigure}[t]{0.45\textwidth}
        \centering
        \resizebox{\linewidth}{!}
        {
        \hspace{12pt}
        \begin{tikzpicture}[every node/.style={draw,circle,minimum size=1.5mm,inner sep=0pt,outer sep=0pt,fill=black}]
            \draw[thick] (0,0) circle (2) ;
            \begin{scope}[rotate=15]
                \path[draw,dashed,thick,blue] (2,0) arc [start angle=0,
                            end angle=180,
                            x radius=2cm,
                            y radius=1cm] ;
                \path[draw, thick, blue, name path global=H1] (-2,0) arc [start angle=180,
                        end angle=360,
                        x radius=2cm,
                        y radius=1cm] ;
            \end{scope}
            \begin{scope}[rotate=165]
                \path[] (2,0) arc 
                [start angle=0,
                            end angle=180,
                            x radius=2cm,
                            y radius=1cm] ;
                \path[name path global=H2] (-2,0) arc [start angle=180,
                        end angle=360,
                        x radius=2cm,
                        y radius=1cm] ;
            \end{scope}
            \node [label=left :$\psi_{0}$,opacity=0.75] at (-1.92,-.55) {};
            \node [label=above :$\psi_{\widehat{\pi}}$] at (1.2,1.6) {};  
            \path [name intersections={of=H1 and H2,by=psi1}];
            \node [label=left :$\psi_1$, blue] at (psi1) {};
        \end{tikzpicture}
        }
        \caption{The vector $\psi_0$ is used to define the subspace $\hH_1$.  The unit vectors in $\hH_1$ define a lower dimensional sphere in $\hH$ (in blue).  The vector $\psi_1$ is the vector in that sphere that is closest to $\psi_{\widehat{\pi}}$.}
    \end{subfigure}
    \vskip\baselineskip
    \begin{subfigure}[t]{0.45\textwidth}
        \centering
        \resizebox{\linewidth}{!}
        {
        \hspace{12pt}
        \begin{tikzpicture}[every node/.style={draw,circle,minimum size=1.5mm,inner sep=0pt,outer sep=0pt,fill=black}]
            \draw[thick] (0,0) circle (2) ;
            \begin{scope}[rotate=15]
                \path[draw,dashed,thin,,opacity=0.75] (2,0) arc [start angle=0,
                            end angle=180,
                            x radius=2cm,
                            y radius=1cm] ;
                \path[draw, thin, opacity=0.75, name path global=H1] (-2,0) arc [start angle=180,
                        end angle=360,
                        x radius=2cm,
                        y radius=1cm] ;
            \end{scope}
            \begin{scope}[rotate=165]
                \path[draw,dashed,thick, blue] (2,0) arc 
                [start angle=0,
                            end angle=180,
                            x radius=2cm,
                            y radius=1cm] ;
                \path[draw,thick, blue, name path global=H2] (-2,0) arc [start angle=180,
                        end angle=360,
                        x radius=2cm,
                        y radius=1cm] ;
            \end{scope}
            \begin{scope}[rotate=-270]
                \path[] (2,0) arc [start angle=0,
                            end angle=180,
                            x radius=2cm,
                            y radius=1cm] ;
                \path[name path global=H3] (-2,0) arc [start angle=180,
                        end angle=360,
                        x radius=2cm,
                        y radius=1cm] ;
            \end{scope}
            \node [label=left :$\psi_{0}$,opacity=0.5] at (-1.92,-.55) {};
            \node [label=above :$\psi_{\widehat{\pi}}$] at (1.2,1.6) {};  
            \path [name intersections={of=H1 and H2,by=psi1}];
            \node [label=left :$\psi_1$,opacity=0.75] at (psi1) {};
            \path [name intersections={of=H2 and H3,by=psi2}];
            \node [label=left:$\psi_2$, blue] at (psi2) {};
        \end{tikzpicture}
        }
        \caption{The vector $\psi_1$ is used to define the subspace $\hH_2$.  The unit sphere in $\hH_2$ (in blue) contains $\psi_1$ but does not contain $\psi_0$.  The vector $\psi_2$ is the unit vector in $\hH_2$ closest to $\psi_{\widehat{\pi}}$. }
    \end{subfigure}
    \hfill
    \begin{subfigure}[t]{0.45\textwidth}
        \centering
        \resizebox{\linewidth}{!}
        {
        \hspace{12pt}
        \begin{tikzpicture}[every node/.style={draw,circle,minimum size=1.5mm,inner sep=0pt,outer sep=0pt,fill=black}]
            \draw[thick] (0,0) circle (2) ;
            \begin{scope}[rotate=15]
                \path[draw,dashed,line width=.1mm,opacity=0.25] (2,0) arc [start angle=0,
                            end angle=180,
                            x radius=2cm,
                            y radius=1cm] ;
                \path[draw, line width=.1mm,opacity=0.25,name path global=H1] (-2,0) arc [start angle=180,
                        end angle=360,
                        x radius=2cm,
                        y radius=1cm] ;
            \end{scope}
            \begin{scope}[rotate=165]
                \path[draw,dashed,thin,opacity=0.5] (2,0) arc 
                [start angle=0,
                            end angle=180,
                            x radius=2cm,
                            y radius=1cm] ;
                \path[draw,thin,opacity=0.5,name path global=H2] (-2,0) arc [start angle=180,
                        end angle=360,
                        x radius=2cm,
                        y radius=1cm] ;
            \end{scope}
            \begin{scope}[rotate=-270]
                \path[draw,dashed,thick, blue] (2,0) arc [start angle=0,
                            end angle=180,
                            x radius=2cm,
                            y radius=1cm] ;
                \path[draw, thick, blue,name path global=H3] (-2,0) arc [start angle=180,
                        end angle=360,
                        x radius=2cm,
                        y radius=1cm] ;
            \end{scope}
            \node [label=left :$\psi_{0}$,opacity=0.25] at (-1.92,-.55) {};
            \node [label=above :$\psi_{\widehat{\pi}}$] at (1.2,1.6) {};  
            \path [name intersections={of=H1 and H2,by=psi1}];
            \node [label=left :$\psi_1$,opacity=0.5] at (psi1) {};
            \path [name intersections={of=H2 and H3,by=psi2}];
            \node [label=left:$\psi_2$,opacity=0.75] at (psi2) {};
            \node [label=left :$\psi_3$, blue] (psi3) at (.63,1.55) {};
        \end{tikzpicture}
        }
        \caption{The vector $\psi_2$ is used to define the subspace $\hH_3$.  The vector $\psi_3$ is the unit vector in $\hH_3$ closest to $\psi_{\widehat{\pi}}$. And so on.}
    \end{subfigure}
    \caption{A bird's eye view of the training dynamics of exact single-site DMRG on the unit sphere.}\label{spheres}

\end{figure}